\documentclass[aps,pra,reprint,showpacs,superscriptaddress,nofootinbib]{revtex4-1}
\usepackage{amsmath,amssymb,amsfonts}
\usepackage{graphicx}
\usepackage{dcolumn}
\usepackage{epstopdf}
\usepackage[T1]{fontenc}
\usepackage{newtxtext}
\usepackage{newtxmath}
\usepackage{subfigure}
\usepackage{float}
\usepackage{bm}

\usepackage[unicode]{hyperref}
\hypersetup{
	unicode=true,          
	a4paper=true,
	plainpages=false,
	pdftitle={Title of PDF},    
	pdfauthor={Author of PDF},     
	pdfsubject={Subject of PDF},   
	colorlinks=true,       
	linkcolor=blue,          
	citecolor=blue,        
	filecolor=blue,      
	urlcolor=blue           
}
\newcommand{\refeq}[1]{Eq.~(\ref{#1})}
\newcommand{\refequation}[1]{Equation~(\ref{#1})}

\newcommand{\refeqsand}[2]{Eqs.~(\ref{#1}) and (\ref{#2})}
\newcommand{\refequationsand}[2]{Equations~(\ref{#1}) and (\ref{#2})}
\newcommand{\reffigsand}[2]{Figs.~\ref{#1} and \ref{#2}}
\newcommand{\reffigs}[4]{Figs.~\ref{#1}(#2) -- \ref{#3}(#4)}
\newcommand{\reffig}[1]{Fig.~\ref{#1}}
\newcommand{\reffigure}[1]{Figure~\ref{#1}}

\DeclareMathAlphabet{\mathcal}{OMS}{cmsy}{m,b}{n,it}

\begin{document}

\title{Fingering instabilities and pattern formation in a two-component dipolar Bose-Einstein condensate}

\date{\today}

\author{Kui-Tian Xi}
\email{xi.99@osu.edu}
\email{kuitian.xi@nyu.edu}
\affiliation{Department of Physics, The Ohio State University, Columbus, Ohio 43210, USA}
\affiliation{New York University Shanghai, 1555 Century Avenue, Pudong, Shanghai 200122, China}

\author{Tim Byrnes}
\affiliation{State Key Laboratory of Precision Spectroscopy, School of Physical and Material Sciences, East China Normal University, Shanghai 200062, China}
\affiliation{New York University Shanghai, 1555 Century Avenue, Pudong, Shanghai 200122, China}
\affiliation{NYU-ECNU Institute of Physics at NYU Shanghai, 3663 Zhongshan Road North, Shanghai 200062, China}
\affiliation{National Institute of Informatics, 2-1-2 Hitotsubashi, Chiyoda-ku, Tokyo 101-8430, Japan}
\affiliation{Department of Physics, New York University, New York, New York 10003, USA}

\author{Hiroki Saito}
\affiliation{Department of Engineering Science, University of Electro-Communications, Tokyo 182-8585, Japan}

\begin{abstract}
We study fingering instabilities and pattern formation at the interface of an oppositely polarized two-component Bose-Einstein condensate with strong dipole-dipole interactions in three dimensions. It is shown that the rotational symmetry is spontaneously broken by fingering instability when the dipole-dipole interactions are strengthened. Frog-shaped and mushroom-shaped patterns emerge during the dynamics due to the dipolar interactions. We also demonstrate the spontaneous density modulation and domain growth of a two-component dipolar BEC in the dynamics. Bogoliubov analyses in the two-dimensional approximation are performed, and the characteristic lengths of the domains are estimated analytically. Patterns resembling those in magnetic classical fluids are modulated when the number ratio of atoms, the trap ratio of the external potential, or tilted polarization with respect to the $ z $ direction is varied.
\end{abstract}

\maketitle

\section{Introduction}\label{sec:intro}

Fingering instabilities (e.g., the Saffman-Taylor instability and Rayleigh-Taylor instability) lead to complicated pattern formations of an interface \cite{instarmp1986,instarmp2009}, and ubiquitously occur at the interface between two fluids of different densities and viscosities, such as water suspended atop oil in gravity \cite{keiser2017}, Hele-Shaw cells between two plates \cite{hscell2012,hscell2015,hscell2017}, biological systems \cite{bio2010}, mushroom clouds from volcanic eruptions and nuclear explosions in the atmosphere, plasma fusion \cite{plasma2000-1,plasma2000-2}, supernova explosions \cite{supernova2002,supernova2013}, and the Crab Nebula \cite{crab2008}. In addition, domains of magnetic fluids are also known to undergo fingering instability and form complex labyrinthine patterns \cite{magfluid1975,magfluid1994}.  In resemblance to the classical fluids, a system of two superfluids can also exhibit a Rayleigh-Taylor interfacial instability, for instance, a system of a two-component Bose-Einstein condensate (BEC) \cite{fibec1,fibec2,fibec3,fibec4,fibec5}. A two-component BEC of an atomic gas with strong dipole-dipole interactions (DDIs) exhibits similarities with those in classical magnetic fluids, such as hexagonal, solitonlike, and labyrinthine pattern formations \cite{saito2009}. 

Recently, a dipolar BEC with strong DDIs was shown to exhibit self-organized crystallization, which is observed in the quench dynamics \cite{kadau2016}, and leads to novel self-bound droplet states of superfluids \cite{schmitt2016,barbut2016}. These emerging phenomena have inspired many theoretical studies to clarify the implicit new physics behind stabilization against collapse and droplet pattern formation \cite{xi2016,bisset2016_1,blakie2016,saito2016,wachtler2016_1,wachtler2016_2,bisset2016_2,baillie2016}. 

In this paper, we consider a two-component BEC with strong DDIs in which the dipole moments of the two components are polarized oppositely. By modulating the strength of DDI, we demonstrate that spontaneous symmetry breaking and fingering instabilities occur at the interface between the two components, which leads to frog-shaped and mushroom-shaped pattern formations. The Bogoliubov dispersion relation corresponding to the fingering instability is analyzed in a two-dimensional approximation. Spontaneous density modulation and domain growth are also shown in the dynamics, along with analytical estimations of the characteristic lengths of the dipolar domains. We observe the formation of droplet patterns when the population imbalance in two components becomes large.  Furthermore, a labyrinthine pattern grows as the trap ratio increases and a stripe phase occurs as the tilted angle increases.

This paper is organized as follows. In Sec. \ref{sec:formulation}, we formulate the problem. In Sec. \ref{sec:instability}, we demonstrate the symmetry-breaking fingering instability (Sec. \ref{subsec:fingering}) and domain dynamics of the system (Sec. \ref{subsec:domain}), where a Bogoliubov analysis and an estimation of the characteristic lengths of the dipolar domains are performed. In Sec. \ref{sec:pattern}, we show stationary pattern formation as the number ratio of atoms (Sec. \ref{subsec:number}), trap ratio (Sec. \ref{subsec:trap}), and tilted polarization (Sec. \ref{subsec:tilt}) are varied. In Sec. \ref{sec:conclusion}, we provide the conclusions of our study.

\section{Formulation of the problem}\label{sec:formulation}

A two-component BEC with DDI at zero temperature is described by the macroscopic wave functions $ \Psi_{1} \left( \bm{r}, t \right)  $ and $ \Psi_{2} \left( \bm{r}, t\right) $ in the mean-field regime, which obey the nonlocal Gross-Pitaevskii (GP) equations given by 
\begin{eqnarray}
	i \hbar \frac{\partial}{\partial t} \Psi_{i} \left(\bm{r}\right)
	&=& \Bigg[ -\frac{\hbar^2}{2m} \nabla^2 + V \left( \bm{r} \right) + \sum_{j = 1}^{2} g_{ij} \big| \Psi_{j}\left(\bm{r}\right) \big|^2 \nonumber\\
	& & + \sum_{j = 1}^{2} \int U_{ij} \left( \bm{r} - \bm{r'} \right) \big| \Psi_{j} \left( \bm{r'} \right) \big|^2 d \bm{r'} \Bigg] \Psi_{i}\left(\bm{r}\right),\quad \label{gp}
\end{eqnarray}
where the atomic mass $ m $ is assumed to be the same for both components. The two-body coupling constants are $ g_{ij} = 4 \pi \hbar^2 a_{ij} / m $, where $ a_{ij} $ is the $ s $-wave scattering length between components $ i $ and $ j $. The macroscopic wave function $ \Psi_{i} $ is normalized as $ \int |\Psi_{i}|^2 d \bm{r} = N_{i} $, where $ N_{i} $ is the number of atoms in component $ i $. The DDI has the form 
\begin{equation}\label{ddi}
	U_{ij} \left( \bm{r} \right) = \frac{\mu_{0} \mu_{i} \mu_{j} \left[ 1 - 3 \left( \hat{d} \cdot \hat{r} \right)^{2} \right]}{4 \pi r^{3}},
\end{equation}
where $ \mu_{0} $ is the magnetic permeability of vacuum, $ \mu_{i} $ is the magnetic dipole moment of component $ i $, $ \hat{d} = \cos \alpha \hat{z} + \sin \alpha \hat{x} $ is the direction of the dipole polarization with $\alpha$ being the angle between $ \hat{d} $ and the $ z $ axis, $ \bm{r} $ is the vector between the two dipoles, and $ \hat{r} = \bm{r} / |\bm{r}| $. In our calculations, the atoms are trapped in the same axisymmetric harmonic potential $ V = m[ \omega_{\perp}^{2} ( x^2 + y^2 ) + \omega_{z}^{2} z^2 ] / 2 $, where $ \omega_{\perp} $ and $ \omega_{z} $ are the radial and axial trap frequencies, respectively.

In our numerical simulations, we solve the three-dimensional nonlocal GP equations in \refeq{gp} using the pseudospectral method with a fast Fourier transform, where the dipolar interaction terms in \refeq{ddi} are calculated using the convolution theorem. We take $ ^{52} $Cr as the atoms in the two-component system, with $ \mu_{1} = 6 \mu_{B} $ and $ \mu_{2} = - 6 \mu_{B} $, where $ \mu_{B} $ is the Bohr magneton. Experimentally, such a two-component dipolar BEC system can be implemented by using $ ^{7}S_{3} $ $ m_{J} = -3 $ and $ +3 $ states of $ ^{52} $Cr \cite{crbec2005,crbec2006}, and the $ s $-wave scattering lengths $ a_{ij} $ can be modulated by using Feshbach resonance \cite{lahaye2007}.

\section{Instabilities and dynamics}\label{sec:instability}

\subsection{Fingering instabilities}\label{subsec:fingering}

\begin{figure*}[t]
	\begin{center}
		\includegraphics[width=1\textwidth]{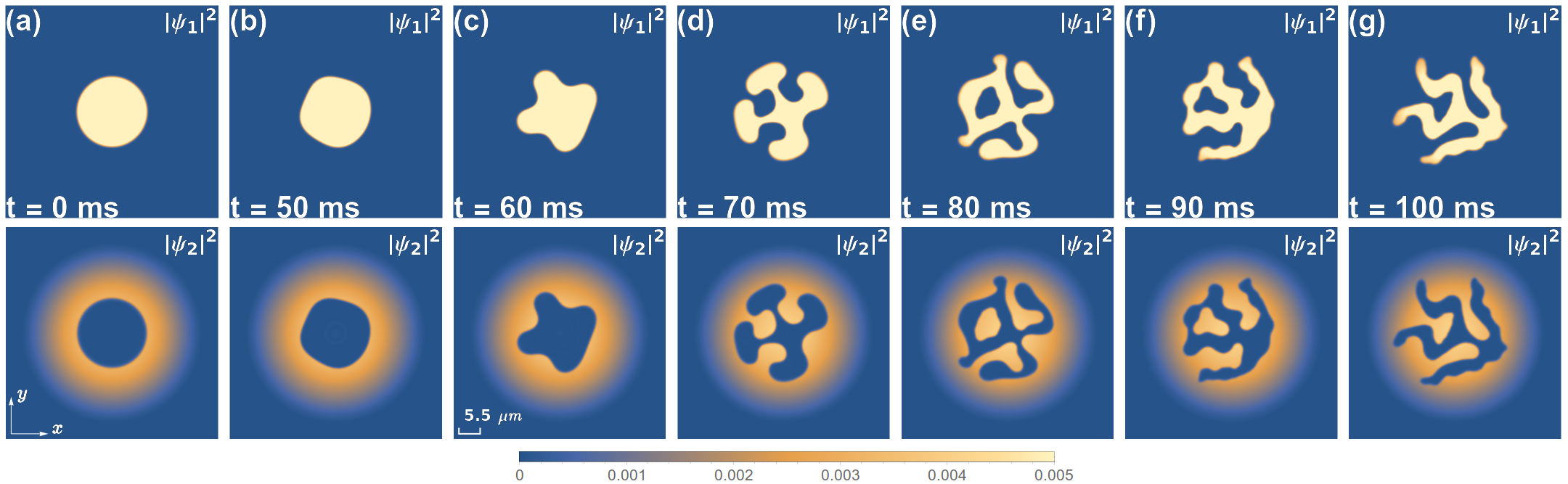}
	\end{center}\vspace{-0.5cm}
	\caption{\label{fig:gamma018} (Color online) Column density profiles $ |\psi_{1}|^{2} $ and $ |\psi_{2}|^{2} $ for dynamical states at $ \gamma = 0.18 $ starting with the non-dipolar ground state. The other parameters are $ a_{11} = 100 a_{B} $, $ a_{22} = 1.1 a_{11} $, $ a_{12} = 1.2 a_{11} $, $ \left( \omega_{\perp}, \omega_{z} \right) = 2 \pi \times \left( 100, 800 \right) $ Hz, $ N_{1} = 1 \times 10^6 $, $ N_{2} = 2 \times 10^6 $, $ \mu_{1} = 6\mu_{B} $, and $ \mu_{2} = -6\mu_{B} $. The field of view is $ 55.66 \times 55.66 $ $ \mu{\rm m} $. The gauge in (c) represents the most unstable wavelength predicted from \refeq{varpi}. The unit for the density plot is $ N_{2} a_{\perp}^{-2} $. See Supplemental Material for a movie of the dynamics \cite{movie}.}
\end{figure*}

\begin{figure*}[t]
	\begin{center}
		\includegraphics[width=1\textwidth]{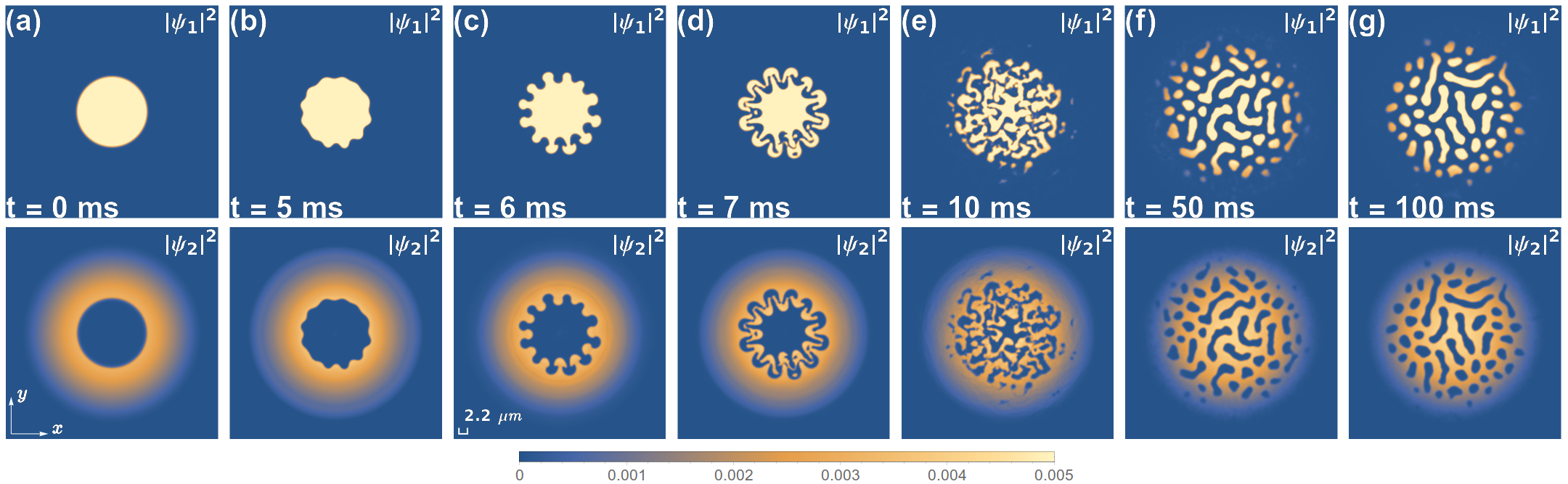}
	\end{center}\vspace{-0.5cm}
	\caption{\label{fig:gamma1} (Color online) Column density profiles $ |\psi_{1}|^{2} $ and $ |\psi_{2}|^{2} $ for dynamical states at $ \gamma = 1 $ starting with the non-dipolar ground state. The other parameters are the same as in \reffig{fig:gamma018}. The field of view is $ 55.66 \times 55.66 $ $ \mu{\rm m} $. The gauge in (c) represents the most unstable wavelength predicted from \refeq{varpi}. The unit for the density plot is $ N_{2} a_{\perp}^{-2} $. See Supplemental Material for a movie of the dynamics \cite{movie}.}
\end{figure*}

\begin{figure*}[t]
	\begin{center}
		\includegraphics[width=1\textwidth]{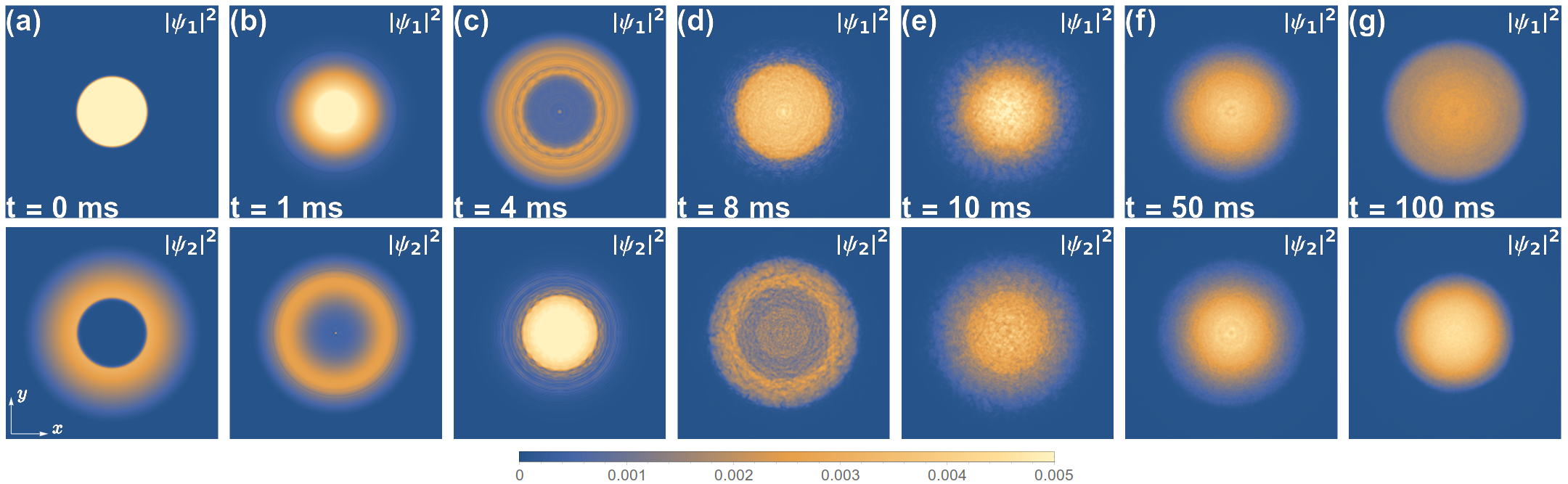}
	\end{center}\vspace{-0.5cm}
	\caption{\label{fig:gamma0} (Color online) Column density profiles $ |\psi_{1}|^{2} $ and $ |\psi_{2}|^{2} $ for $ \gamma = 0 $ starting with the ground state for $ a_{11} = a_{22} = a_{12} = 100 a_{B} $. The other parameters are the same as in \reffig{fig:gamma018}, except that the scattering lengths are changed to $ a_{11} = a_{22} = 100 a_{B} $ and $ a_{12} = 0.5 a_{11} $ at $ t = 0 $. The field of view is $ 55.66 \times 55.66 $ $ \mu{\rm m} $. The unit for the density plot is $ N_{2} a_{\perp}^{-2} $. See Supplemental Material for a movie of the dynamics \cite{movie}.}
\end{figure*}

We now investigate the fingering instabilities at the interface between two components with different strengths of dipolar interactions with $ \alpha = 0 $, i.e., there is no tilted polarization in this case. To investigate the dependence of the DDI strength on the dynamics, we introduce the strength coefficient $ \gamma $ of the DDI as 
\begin{equation}\label{ddi2}
	U_{ij} \left( \bm{r} \right) = \gamma \frac{\mu_{0} \mu_{i} \mu_{j} \left( 1 - 3 \cos^{2} \theta \right)}{4 \pi r^3},
\end{equation}
where $ \cos \theta = \hat{d} \cdot \hat{r} = \hat{z} \cdot \hat{r} $. Experimentally, the coefficient $\gamma$ can be tuned by fast rotation of the magnetic field \cite{gamma}. We initially prepare stationary states without DDI, which become metastable states with unstable interfaces in the presence of DDI. The DDI is then suddenly introduced at $t = 0$, followed by time evolution. The parameters are taken to be $ a_{11} = 100 a_{B} $, $ a_{22} = 1.1 a_{11} $, $ a_{12} = 1.2 a_{11} $, $ \left( \omega_{\perp}, \omega_{z} \right) = 2 \pi \times \left( 100, 800 \right) $ Hz, and $ N_{1} = 1 \times 10^6 $, $ N_{2} = 2 \times 10^6 $. These scattering lengths satisfy the phase-separation condition, $g_{11} g_{22} < g_{12}^{2} $, and the ground state without DDI has a circular interface between the two components.

The column density profiles obtained by the integration of the densities along the $ z $ direction are
\begin{equation}
\big| \psi_{i} \big|^{2} = \int \big| \Psi_{i} \big|^{2} dz, 
\end{equation}\label{column}
where $ i = 1 $, $ 2 $.
These column densities are shown in \reffigsand{fig:gamma018}{fig:gamma1} for $ \gamma = 0.18 $ and $ 1 $, respectively. In the absence of the DDI, the immiscible states of the two components exhibit rotational symmetry, as expected, which are shown in \reffig{fig:gamma018}(a) and \reffig{fig:gamma1}(a). After the DDI is induced, the rotational symmetry of the circular interface is spontaneously broken, leading to a fingering instability at the interface between the two components, which is similar to the interface behavior of classical dipolar fluids \cite{magfluid1994}. We find from \reffigsand{fig:gamma018}{fig:gamma1} that the system exhibits spontaneous rotational symmetry breaking with a smaller length scale for a larger DDI. For $ \gamma = 0.18 $, a frog-shaped pattern is formed in component 1 at $ t = 100 $ ms, as shown in \reffig{fig:gamma018}(g). For $ \gamma = 1 $, a mushroom-shaped pattern first appears at the interface, as shown in \reffig{fig:gamma1}(c) and \reffig{fig:gamma1}(d), and then the droplet pattern is formed at a later time, as shown in \reffig{fig:gamma1}(f) and \reffig{fig:gamma1}(g). It is shown that with stronger dipolar interactions, the interface of a two-component BEC becomes more unstable.

For comparison, we then investigate the case in the absence of DDI, in which the scattering lengths are suddenly changed from immiscible to miscible, $ g_{11} g_{22} > g_{12}^{2} $ at $ t = 0 $. As shown in \reffig{fig:gamma0}, no fingering instability occurs at the interface without DDI (i.e., $ \gamma = 0 $ together with the same values of the other parameters), and the initially phase-separated components mix in the dynamics, which demonstrates that the immiscible-miscible transition has no contribution to the fingering instability at the interface. These results reveal that fingering instability never appears due to the immiscible-miscible transition.

For a deeper understanding of the instability, we perform a Bogoliubov analysis for the system. To simplify the calculation, the dimensionality of the system is reduced to two dimensions (2D). We assume that the $z$ dependence of the wave function is decomposed as 
\begin{equation}
	\Psi_{i} \left( \bm{r}, t \right) = \psi_{i} \left( \bm{\rho}, t \right) \frac{1}{\left( \pi \zeta^{2} \right)^{1/4}} e^{-z^{2} / \left( 2 \zeta^{2} \right)},
\end{equation} 
where $\bm{\rho} = (x, y)$ and $\zeta$ is the width of the density profile in the $z$ direction. The value of $\zeta$ is determined by fitting the Gaussian to the numerically obtained wave function. \refequation{gp} is then rewritten as 
\begin{eqnarray}
	i \hbar \frac{\partial}{\partial t} \psi_{i} &=& \Bigg[ -\frac{\hbar^2}{2m} \nabla_{\bm{\rho}}^2 + V \left( \bm{\rho} \right) + \frac{\hbar\omega_{z}}{4} \left( \frac{a_{z}^{2}}{\zeta^{2}} + \frac{\zeta^{2}}{a_{z}^{2}} \right)  + \sum_{j = 1}^{2} g_{ij}^{2D} \big| \psi_{j} \big|^2 \nonumber\\
	& & + \sum_{j = 1}^{2} \int U_{ij}^{2D} \left( \bm{\rho} - \bm{\rho'} \right) \big| \psi_{j} \left( \bm{\rho'} \right) \big|^2 d \bm{\rho'} \Bigg] \psi_{i}, \label{gp2d}
\end{eqnarray}
where $ V(\rho) = m \omega_{\perp}^2 (x^2 + y^2) / 2 $, $ a_{z} = \sqrt{\hbar / (m \omega_{z})} $ and $ g_{ij}^{2D} = g_{ij} / \small\sqrt{2 \pi \zeta^{2}} $. The 2D dipolar interaction reads \cite{tiltmean}  
\begin{equation}
	U_{ij}^{2D} \left( \bm{\rho} - \bm{\rho'} \right) = \gamma \frac{\mu_{0} \mu_{i} \mu_{j}}{3 \sqrt{2 \pi \zeta^{2}}} \int \frac{d \bm{k}}{\left( 2 \pi \right)^{2}} e^{-i \bm{k} \cdot \left( \bm{\rho} - \bm{\rho'} \right)} h \left( q \right),
\end{equation}
where the dipolar kernel $ h(q) = 2 - 3 \small\sqrt{\pi} q e^{q^{2}} \mbox{erfc} (q) $ with $ q = k \zeta / \small\sqrt{2} $ for polarized spins in the $ z $ direction.

We define $ \phi_{i} (\rho) $ as the stationary state with rotational symmetry, and then weak perturbations of the wave functions can be expressed as,
\begin{eqnarray}
	\psi_{i}\left(\bm{\rho}, t\right) = e^{-i M_{i} t / \hbar} \Big[ \phi_{i}\left(\rho\right) &+& u_{i}\left(\rho\right) e^{iL \varphi} e^{-i\Omega t} \nonumber\\
	&+& v_{i}^{*}\left(\rho\right) e^{-iL \varphi} e^{i\Omega^{*} t} \Big], \label{phi}
\end{eqnarray}
where $ M_{i} $ is the chemical potential of component $ i $, the amplitudes $ u_{i}(\rho) $ and $ v_{i}(\rho) $ are treated as small, $ \varphi = \tan^{-1} (y / x) $, and $ \Omega $ is the frequency of the oscillation. The quantum number $ L $ characterizes the angular momentum of the modes.

Substituting \refeq{phi} into \refeq{gp2d}, and collecting the first-order terms proportional separately to $ e^{\pm i \Omega t} $, we then obtain the following Bogoliubov de Gennes (BdG) equations,
\begin{widetext}
\begin{eqnarray}
	\hbar \Omega u_{i} &=& \Big( H_{0} - M_{i} \Big) u_{i} + \sum_{j=1}^{2} g_{ij}^{2D} \left( \big| \phi_{j} \big|^{2} u_{i} + \phi_{j}^{\ast}\phi_{i} u_{j} + \phi_{j} \phi_{i} v_{j} \right) \nonumber\\
	& & + \int d\bm{\rho}' \sum_{j=1}^{2} U_{ij}^{2D} \Bigg\{ \Big| \phi_{j} \left( \rho' \right) \Big|^{2} u_{i} + \bigg[ \phi_{j}^{\ast} \left( \rho' \right) \phi_{i} \left( \rho \right) u_{j} \left( \rho' \right) + \phi_{j} \left( \rho' \right) \phi_{i} \left( \rho \right) v_{j} \left( \rho' \right) \bigg] e^{iL\varphi'} \Bigg\}, \label{u}\\
	-\hbar \Omega v_{i} &=& \Big( H_{0} - M_{i} \Big) v_{i} + \sum_{j=1}^{2} g_{ij}^{2D} \left( \big| \phi_{j} \big|^{2} v_{i} + \phi_{j}^{\ast}\phi_{i} v_{j} + \phi_{j} \phi_{i} u_{j} \right) \nonumber\\
	& & + \int d\bm{\rho}' \sum_{j=1}^{2} U_{ij}^{2D} \Bigg\{ \Big| \phi_{j} \left( \rho' \right) \Big|^{2} v_{i} + \bigg[ \phi_{j}^{\ast} \left( \rho' \right) \phi_{i} \left( \rho \right) v_{j} \left( \rho' \right) + \phi_{j} \left( \rho' \right) \phi_{i} \left( \rho \right) u_{j} \left( \rho' \right) \bigg] e^{iL\varphi'} \Bigg\}, \label{v}
\end{eqnarray}
\end{widetext}
where $ H_{0} = -\hbar^{2} \nabla^{2} /(2m) + V(\rho) $, and $ i = 1 $, $ 2 $.

\begin{figure}[t]
	\begin{center}
		\includegraphics[width=0.48\textwidth]{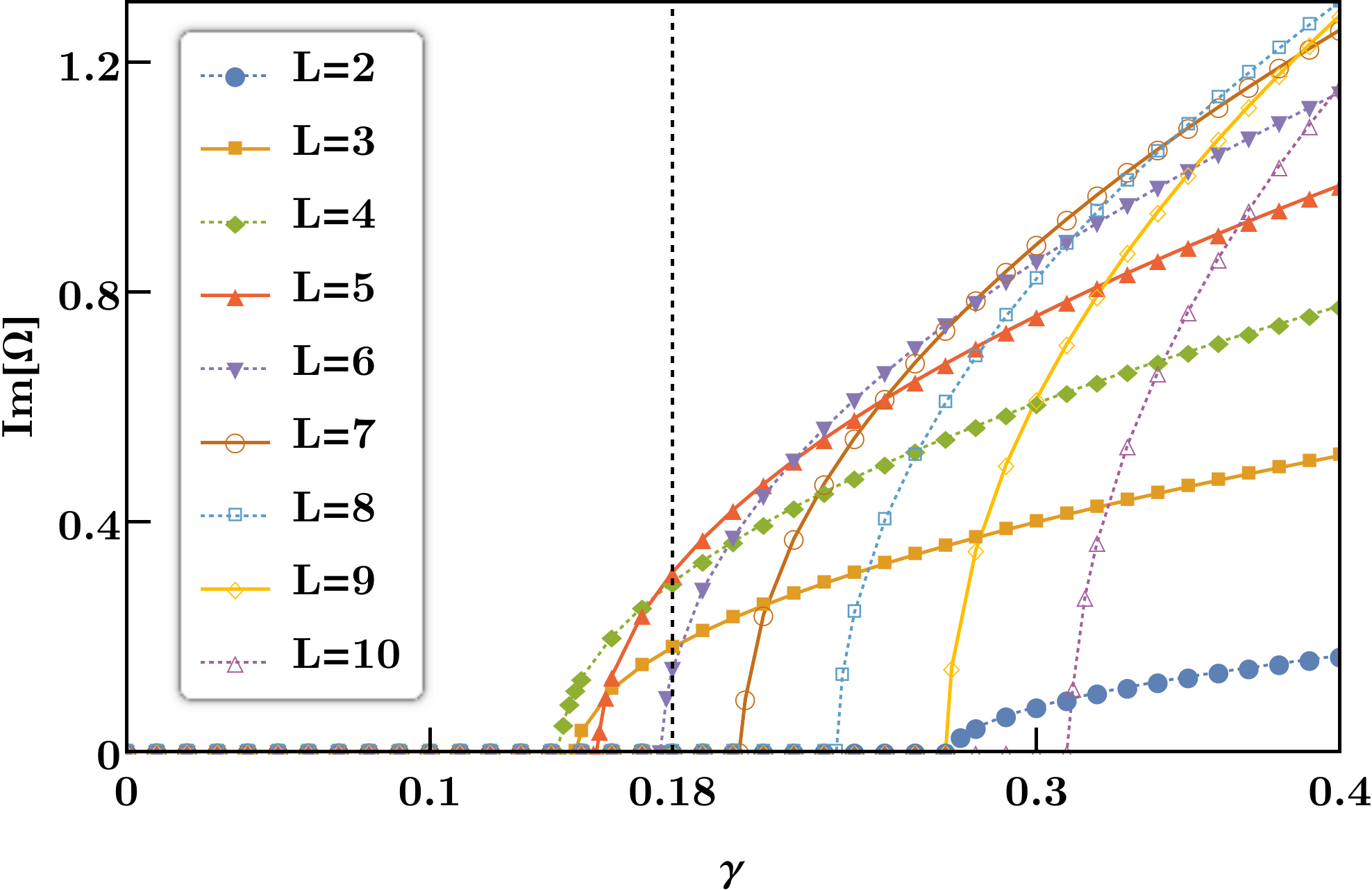}
	\end{center}\vspace{-0.5cm}
	\caption{\label{fig:bogo} (Color online) Imaginary part of Bogoliubov excitation frequency $ {\rm Im}[\Omega] $ as a function of $ \gamma $. The modes for $ 2 \leq L \leq 10 $ are plotted, while $ {\rm Im}[\Omega] = 0 $ for the $ L = 0 $ and $ 1 $ modes. The vertical black dashed line indicates $ \gamma = 0.18 $, corresponding to the parameter in \reffig{fig:gamma018}.}
\end{figure}

We numerically diagonalize \refeqsand{u}{v} to study the instability of the system. If $ \Omega $ has a positive imaginary part, the system exhibits a dynamical instability, and the unstable modes will grow exponentially to the evolution limitation arising from the nonlinearities. The imaginary part of the Bogoliubov excitation frequency $ {\rm Im}[\Omega] $ as a function of dipolar strength coefficient $ \gamma $ is shown in \reffig{fig:bogo} with modes $ L = 2 - 10 $, among which the modes with the largest $ {\rm Im}[\Omega] $ dominate the instability dynamics. For the $ L = 0 $ and $ 1 $ modes (monopole and dipole modes), $ {\rm Im}[\Omega] = 0 $ and these modes are stable. Each mode in $ L = 2 - 10 $ has its critical strength of dipolar interaction, above which $ {\rm Im}[\Omega] > 0 $  and monotonically rises. When $ \gamma = 0.18 $ (the vertical black dashed line in \reffig{fig:bogo}) corresponding to \reffig{fig:gamma018}, the interface is unstable with modes $ L = 3 - 6 $, in which the modes of $ L = 4 $ and $ 5 $ with the relatively largest $ {\rm Im}[\Omega] $ are predominant in the instability dynamics. We thus see that the results shown in \reffig{fig:gamma018} are in agreement with the Bogoliubov analysis.

We then analyze the fingering instability at the interface induced by the DDI. The DDI energy of the system in the 2D approximation is given by 
\begin{eqnarray}
	E_{dd} &=& \frac{g_{dd}}{2} \int d\bm{\rho} d\bm{\rho}' \int \frac{d \bm{k}}{\left( 2\pi \right)^{2}} e^{-i \bm{k} \cdot \left( \bm{\rho} - \bm{\rho}' \right)} h\left( q \right) \nonumber\\
	& & \times \Big[ n_{1} \left( \bm{\rho} \right) - n_{2} \left( \bm{\rho} \right) \Big] \Big[ n_{1} \left( \bm{\rho}' \right) - n_{2} \left( \bm{\rho}' \right) \Big],
\end{eqnarray}
where $ g_{dd} = \gamma \mu_{0} (6\mu_{B})^2 / \big( 3 \small\sqrt{2 \pi \zeta^{2}} \big) $, and $ n_{i} =|\psi_{i}|^{2} $ is the 2D density for component $ i $. For simplicity, we assume that components $ 1 $ and $ 2 $ have uniform distributions in $ y \lesssim 0 $ and $ y \gtrsim 0 $, respectively. The interface is located at $ y \simeq 0 $, and the position of the interface is defined by 
\begin{equation}\label{eta}
	y = \eta \left( x, t \right).
\end{equation}
A smoothed step function is introduced as 
\begin{equation}\label{fy}
	f \left( y \right) = \left\{ \begin{array}{ll}
		0 & \mbox{if $ y \rightarrow -\infty $};\\
		1 & \mbox{if $ y \rightarrow +\infty $},
	\end{array} \right.
\end{equation}
and $ f'(y) > 0 $ is localized at $ y \simeq 0 $. Using \refeqsand{eta}{fy}, the densities of the two components can be expressed as 
\begin{eqnarray}
	n_{1} \left( \bm{\rho}, t \right) &=& n_{1}^{0} f \big[ -y + \eta \left( x, t \right) \big], \\
	n_{2} \left( \bm{\rho}, t \right) &=& n_{2}^{0} f \big[ y - \eta \left( x, t \right) \big].
\end{eqnarray}
We assume that $ \eta $ is small and the densities can be expanded as 
\begin{eqnarray}
	n_{1} \left( \bm{\rho}, t \right) &=& n_{1}^{0} f \left( -y \right) + n_{1}^{0} f' \left( y \right) \eta \left( x, t \right), \\
	n_{2} \left( \bm{\rho}, t \right) &=& n_{2}^{0} f \left( y \right) - n_{2}^{0} f' \left( y \right) \eta \left( x, t \right),
\end{eqnarray}
where we assume $ f' \left( -y \right) = f' \left( y \right) $.
By neglecting the terms that are independent of $ \eta $, we obtain the pressure difference at the interface as 
\begin{eqnarray}
	\frac{\delta E_{dd}}{\delta \eta \left( x, t \right)} &=& g_{dd} \left( n_{1}^{0} + n_{2}^{0} \right)^{2} \int dx'  \int \frac{d \bm{k}}{\left( 2 \pi \right)^{2}} e^{-i k_{x} \left( x' - x \right)} \nonumber\\
	& & \times h\left( q \right) g^{2}\left( k_{y} \right) \eta\left( x', t \right),
\end{eqnarray}
where 
\begin{equation}
	g\left( k_{y} \right) = \int dy f'\left(y\right) e^{-ik_{y} y}.
\end{equation}
The function $ f'\left(y\right) $ is assumed to be Gaussian, 
\begin{equation}\label{fyd}
	f'\left(y\right) = \frac{e^{-y^{2} / l_{y}^{2}}}{\sqrt{\pi} l_{y}},
\end{equation}
which gives 
\begin{equation}
	g\left(k_{y}\right) = e^{-k_{y}^{2} l_{y}^{2} / 4},
\end{equation}
where $ l_{y} = \sqrt{ \hbar / ( m \omega_{y} ) } $. 
Assuming that $ \eta \left(x, t\right) \propto \sin \left(\kappa x - \varpi t\right) $, we obtain 
\begin{eqnarray}
	\frac{\delta E_{dd}}{\delta \eta\left(x, t\right)} &=& g_{dd} \left(n_{1}^{0} + n_{2}^{0}\right)^{2} \eta\left(x, t\right) \nonumber\\
	& & \times \int \frac{dk_{y}}{2\pi} h\left(\frac{\zeta}{\sqrt{2}} \sqrt{\kappa^{2} + k_{y}^{2}}\right) g^{2}\left(k_{y}\right) \\
	&\equiv& F\left(\kappa\right) \eta\left(x, t\right), \label{form}
\end{eqnarray}
which corresponds to the pressure acting on the interface. In the Rayleigh-Taylor instability at the interface between heavy and light fluids in the gravity, the pressure acting on the interface is given by $ g (\rho_{\rm heavy} - \rho_{\rm light}) \eta $. Thus, $ F(\kappa) $ in \refeq{form} plays the role of the difference in the gravitational force on the unit volume of heavy and light fluids in the classical Rayleigh-Taylor instability. According to the Rayleigh-Taylor dispersion relation of the interface mode \cite{instabook}, we then obtain 
\begin{equation}
	\varpi = \sqrt{ \frac{ F\left(\kappa\right) \kappa + \sigma \kappa^{3} }{ m\left(n_{1}^{0} + n_{2}^{0} \right) } }, \label{varpi}
\end{equation}
where $ \sigma $ is the 2D interface tension coefficient. 
When $ F\left(\kappa\right) \kappa + \sigma \kappa^{3} < 0 $, the interface mode with wave number $ \kappa $ is dynamically unstable. When the healing lengths in the two components are almost the same, $ \xi_{1} \simeq \xi_{2} \equiv \xi $, the interface tension coefficient in Ref. \cite{tension} is approximated as $ \sigma \simeq g_{11}^{2D} n_{1}^{0} \xi \sqrt{K-1} $, where $ K = a_{12} / \sqrt{a_{11} a_{22}} $. Using the parameters in \reffig{fig:gamma018} and fitting the numerically obtained interface profile to \refeq{fyd}, we obtain $ \sigma \simeq 0.015 \hbar \omega_{\perp} / a_{\perp} $, and the ratio is $ \zeta / a_{\perp} \simeq 1.3 $. Then the most unstable wavelength is estimated to be $ 2\pi/\kappa \simeq 5.5$ $\mu{\rm m}$ for $ \gamma = 0.18 $ and $ 2\pi/\kappa \simeq 2.2 $ $ \mu{\rm m} $ for $ \gamma = 1 $, which are in qualitative agreement with those in \reffigsand{fig:gamma018}{fig:gamma1}. For a more accurate estimation of the unstable wave number, we must consider the circular geometry of the system. Moreover, the interface tension is modified by the DDI, which should be taken into account.

\subsection{Domain dynamics}\label{subsec:domain}

\begin{figure*}[t]
	\begin{center}
		\includegraphics[width=1\textwidth]{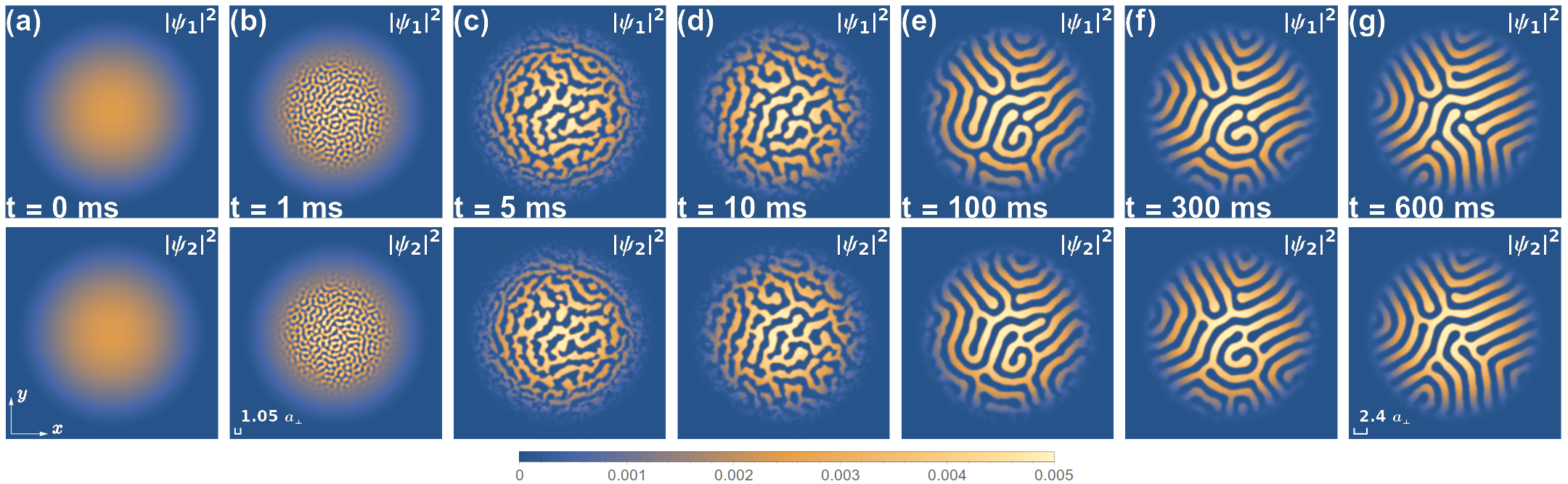}
	\end{center}\vspace{-0.5cm}
	\caption{\label{fig:dynamics} (Color online) Column density profiles $ |\psi_{1}|^{2} $ and $ |\psi_{2}|^{2} $ for dynamical states with different times. The parameters are $ a_{11} = a_{22} = a_{12} = 100 a_{B} $, $ \left( \omega_{\perp}, \omega_{z} \right) = 2 \pi \times \left( 100, 800 \right) $ Hz, $ \gamma = 1 $, $ N_{1} = N_{2} = 2 \times 10^6 $, $ \mu_{1} = 6 \mu_{B} $, and $ \mu_{2} = -6 \mu_{B} $. The field of view is $ 55.66 \times 55.66 $ $ \mu{\rm m} $. The gauges in (c) and (g) represent the wavelengths predicted from \refeqsand{omega2}{E}, respectively. The unit for the density plot is $ N_{1} a_{\perp}^{-2} $. See Supplemental Material for a movie of the dynamics \cite{movie}.}
\end{figure*}

\begin{figure}[t]
	\begin{center}
		\includegraphics[width=0.48\textwidth]{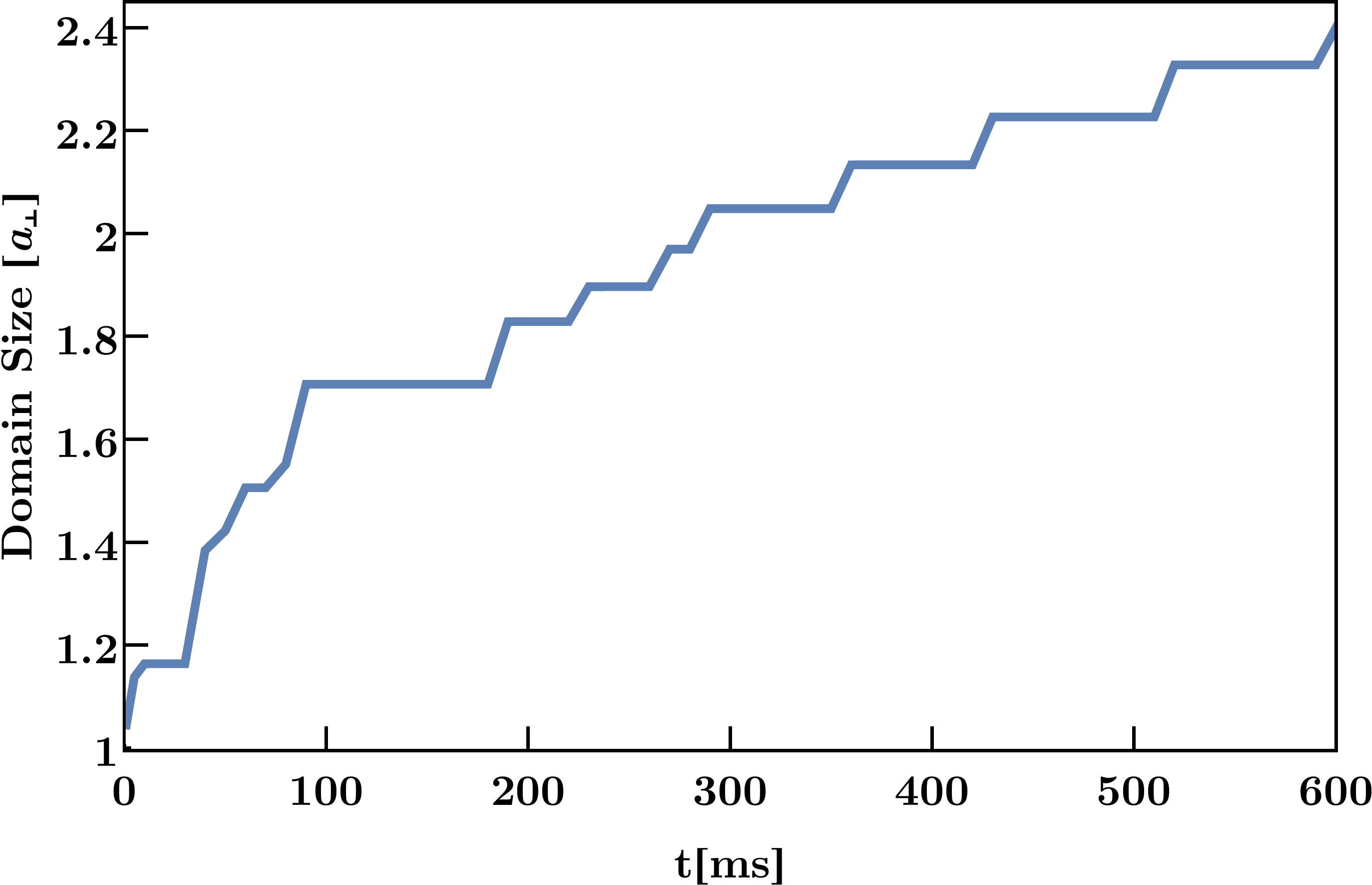}
	\end{center}\vspace{-0.5cm}
	\caption{\label{fig:size} (Color online) Time evolution of the average domain size in the dynamics of \reffig{fig:dynamics}, which is obtained from the Fourier transform of the wave function at each time.}
\end{figure}

We now investigate the domain dynamics for another initial condition. The initial state is the stationary state $ \Psi_{0} $ of a single-component dipolar BEC with the parameters taken as $ \gamma = 1 $, $ N = 4 \times 10^{6} $, $ a = 100 a_{B} $, $ \alpha = 0 $, and $ \left( \omega_{\perp}, \omega_{z} \right) = 2 \pi \times \left( 100, 800 \right) $ Hz. Then the stationary state is split into two components with opposite polarization as $\Psi_{1} = \Psi_{2} = \Psi_{0} / \sqrt{2}$. During the real-time evolution of the two components, we take $ \gamma = 1 $, $ a_{11} = a_{22} = a_{12} = 100 a_{B} $, $ \left( \omega_{\perp}, \omega_{z} \right) = 2 \pi \times \left( 100, 800 \right) $ Hz, and $ N_{1} = N_{2} = 2 \times 10^6 $. In this case, the $ s $-wave interactions have no contribution to the phase separation of the system because of the equal values of $ s $-wave scattering lengths.

The results of the dynamic density modulation in a two-component dipolar BEC are shown in \reffig{fig:dynamics}. At the initial time, the phase separation is induced by the DDI, which leads to a complicated domain structure with a large kinetic energy [\reffig{fig:dynamics}(b)]. The dipolar domains of the system then grow with time [\reffigs{fig:dynamics}{b}{fig:dynamics}{e}], and the domain structure is gradually rearranged at later times [\reffigs{fig:dynamics}{e}{fig:dynamics}{g}]. The domain size in the dynamics is shown as a function of time in \reffig{fig:size}. These results demonstrate that the growth of the domain size is similar to the recent results regarding the hydrodynamics in the magnetic domains of a spin-1 ferromagnetic BEC \cite{domain2011}.

To obtain a deeper understanding of the domain dynamics, we now perform an analytical estimation of the characteristic lengths both at the beginning and at longer times of the dipolar domain formation process in the 2D approximation. At the beginning of the formation of the dipolar domains, the characteristic lengths are estimated by using a Bogoliubov analysis of a uniform system. We define $ g_{11}^{2D} = g_{22}^{2D} = g $, and write the wave functions as 
\begin{equation}
\psi_{i} = e^{-i M t / \hbar} \Big[ \sqrt{n / 2}  + u_{i} e^{i \bm{k} \cdot \bm{\rho}} e^{-i\Omega t} + v_{i}^{*} e^{-i \bm{k} \cdot \bm{\rho}} e^{i\Omega^{*} t} \Big], \label{phidyna}
\end{equation}
where $ n $ is the 2D total density and $ M = n ( g + g_{12}^{2D}) / 2 $. Substituting \refeq{phidyna} into \refeq{gp2d} and diagonalizing the Bogoliubov matrix, we obtain 
\begin{eqnarray}
\hbar \Omega &=& \sqrt{\epsilon_{k} \Big[\epsilon_{k} + \left( g + g_{12}^{2D} \right) n \Big] }, \label{omega1} \\
\hbar \Omega &=& \sqrt{\epsilon_{k} \bigg\{ \epsilon_{k} + \Big[ g - g_{12}^{2D} + 2 g_{dd} h \left( q \right) \Big] n \bigg\} }, \label{omega2}
\end{eqnarray}
where $ \epsilon_{k} = \hbar^{2} k^{2} / (2m) $. \refequationsand{omega1}{omega2} correspond to the eigenfrequencies of the density and spin waves, respectively. If $ \Omega $ is complex, the corresponding mode grows exponentially, and the system is dynamically unstable. Therefore, \refequation{omega2} indicates that the phase separation can occur due to the DDI even when the two components are miscible ($g > g_{12}$) using only $ s $-wave scattering. Substituting $ g_{12}^{2D} = g $ and the parameters used in \reffig{fig:dynamics} into \refeq{omega2}, the most unstable mode, i.e., the wavelength that maximizes the imaginary part of \refeq{omega2}, is $ 2\pi / k_{u} = 1.05 a_{\perp} $ with $ a_{\perp} = \sqrt{\hbar / (m \omega_{\perp})} $, which corresponds to the characteristic wave number for the phase separation at the beginning of the time evolution, as shown in \reffig{fig:dynamics}(b).

At longer times, for the case with a striped labyrinthine pattern in \reffig{fig:dynamics}(g), we apply the trial wave functions of the following form,
\begin{eqnarray}
\psi_{1} &=& \sqrt{n} \sqrt{\frac{1 + \cos K x}{2}}, \label{trial1} \\
\psi_{2} &=& \sqrt{n} \sqrt{\frac{1 - \cos K x}{2}}. \label{trial2}
\end{eqnarray}
The total energy then becomes 
\begin{equation}
\frac{E}{N} = \frac{\hbar^{2} K^{2}}{8m} + \frac{g_{dd} n}{4} h \left(K \zeta / \sqrt{2} \right), \label{E}
\end{equation}
where the $ s $-wave interaction energy is independent of $ K $ and neglected. Substituting the parameters used in \reffig{fig:dynamics} into \refeq{E}, the characteristic domain size at later times	, i.e., the wavelength that minimizes \refeq{E}, is $ 2 \pi / K \simeq 2.4 a_{\perp} $. \reffigure{fig:size} shows the time evolution of the average domain size in the dynamics of \reffig{fig:dynamics}. The domain size is obtained from the average radius of the excitation ring in the Fourier-space density profiles. The initial and final domain sizes in \reffig{fig:size} are in qualitative agreement with $ 2\pi / k_{u} $ and $ 2\pi / K $ estimated above.

\section{Domain patterns in stationary states}\label{sec:pattern}

\subsection{Number ratio}\label{subsec:number}

\begin{figure}[t]
	\begin{center}
		\includegraphics[width=0.485\textwidth]{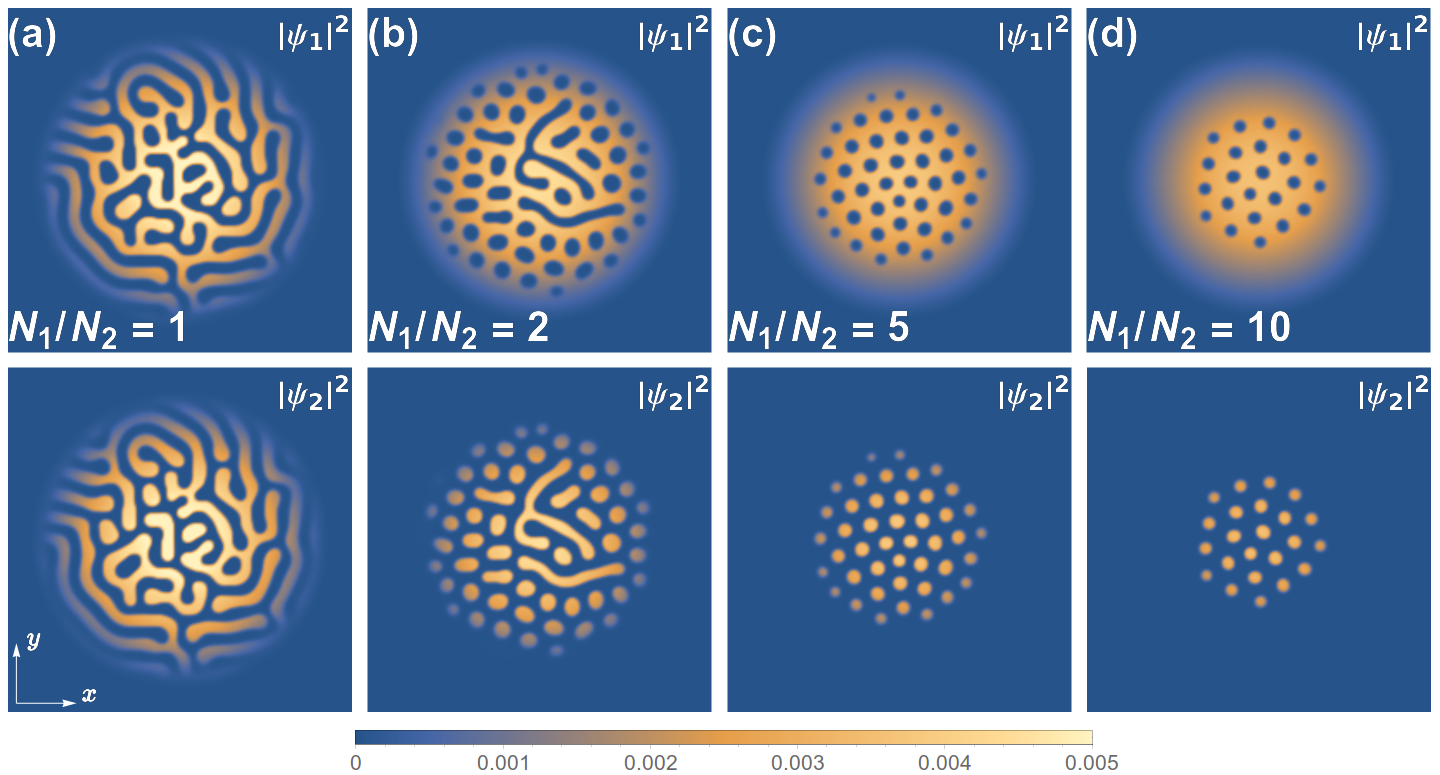}
	\end{center}\vspace{-0.5cm}
	\caption{\label{fig:number} (Color online) Column density profiles $ |\psi_{1}|^{2} $ and $ |\psi_{2}|^{2} $ for stationary states with fixed $ N_{1} $ and different number ratios $ N_{1} / N_{2} $. The other parameters are $ \gamma = 1 $, $ a_{11} = a_{22} = a_{12} = 100 a_{B} $, $ \left( \omega_{\perp}, \omega_{z} \right) = 2 \pi \times \left( 100, 800 \right) $ Hz, $ N_{1} = 2 \times 10^6 $, $ \mu_{1} = 6 \mu_{B} $, and $ \mu_{2} = -6 \mu_{B} $. The field of view is $ 55.66 \times 55.66 $ $ \mu{\rm m} $. The unit for the density plot is $ N_{1} a_{\perp}^{-2} $.}
\end{figure}

Using imaginary-time propagation, we investigate the stationary states with different numbers of atoms $ N_{2} $ in component 2 with fixed $ N_{1} $ in component 1. We take $ a_{11} = a_{22} = a_{12} = 100 a_{B} $, $ \alpha = 0 $, and $ \left( \omega_{\perp}, \omega_{z} \right) = 2 \pi \times \left( 100, 800 \right) $ Hz. The number of atoms in component 1 is fixed to $ N_{1} = 2 \times 10^6 $. In all the following simulations, the dipolar strength coefficient $ \gamma = 1 $ is maintained. The $ s $-wave scattering lengths herein and in the following are set to fulfill the condition of the miscible states. The ratio between the numbers of atoms $ N_{1} / N_{2} $ is changed from $ 1 $ to $ 10 $.

The column density profiles are plotted in \reffig{fig:number}. Phase separation shown in \reffig{fig:number}(a) arises from the opposite polarization of dipole moments in the two components, and rotational symmetry breaking (labyrinthine pattern) appears spontaneously due to the strong DDI. Moreover, a droplet pattern occurs when $ N_{1} / N_{2} > 1 $, where the droplets tend to form a triangular lattice. The crystallization of the dipolar droplets is similar to that observed in Ref. \cite{kadau2016}. The number of droplets decreases as $ N_{2} $ is decreased. For component 2 (bottom row in \reffig{fig:number}), which has a smaller number of atoms, the filament structure splits into droplets and tends to assemble toward the center.  Our results are similar to those in Ref.\cite{ryan2012}, in which only one component is dipolar and the system is strictly two dimensional.

\subsection{Trap ratio}\label{subsec:trap}

\begin{figure}[t]
	\begin{center}
		\includegraphics[width=0.485\textwidth]{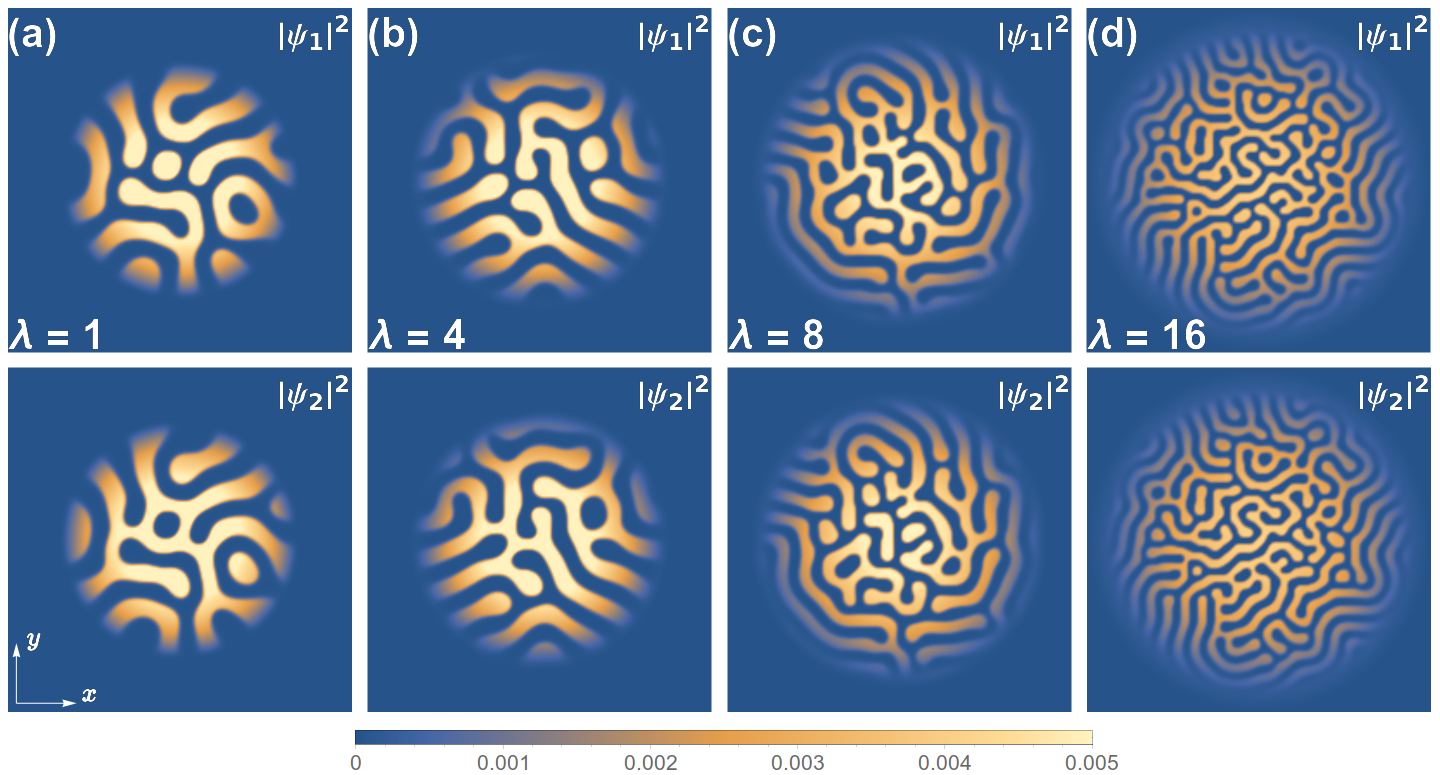}
	\end{center}\vspace{-0.5cm}
	\caption{\label{fig:trap} (Color online) Column density profiles $ |\psi_{1}|^{2} $ and $ |\psi_{2}|^{2} $ for stationary states with different trap ratios $ \lambda $. The other parameters are the same as in \reffig{fig:number}, except that $ N_{2} = 2 \times 10^6 $ and $ \omega_{\perp} = 2 \pi \times 100 $ Hz. The field of view is $ 55.66 \times 55.66 $ $ \mu{\rm m} $. The unit for the density plot is $ N_{1} a_{\perp}^{-2} $.}
\end{figure}

We then investigate the trap geometry effect on the pattern formation in the two-component dipolar BEC. The number of atoms in the two components is the same, $ N_{1} = N_{2} = 2 \times 10^{6} $, $ \alpha = 0 $, and $ a_{11} = a_{22} = a_{12} = 100 a_{B} $ are taken. The radial trap frequency is fixed as $ \omega_{\perp} = 2 \pi \times 100 $ Hz, while the trap ratio defined by $ \lambda = \omega_{z} / \omega_{\perp} $ is changed from $ 1 $ to $ 16 $. 

The column density profiles with $ \lambda = 1 $, $ 4 $, $ 8 $, and $ 16 $ are plotted in \reffig{fig:trap}. As the trap ratio increases with the fixed radial trap potential, the axial trap potential is enhanced, and the atoms in the trap tend to be expanded toward the outside along the radial directions, since the density increases with an increase in the axial confinement. Thus, a more complicated labyrinthine pattern is formed for larger $\lambda$ with the radial trap frequency being fixed. The pattern in \reffig{fig:trap}(c) is the same as that in \reffig{fig:number}(a) due to the same parameter settings. In the imaginary-time evolution, small noise is introduced in the initial state, and the complicated labyrinthine structures obtained after convergence depend on the initial noise. Therefore, a lot of similar structures are almost degenerate near the ground state, and each of them is obtained depending on the initial noise.

\subsection{Tilted polarization}\label{subsec:tilt}

\begin{figure}[t]
	\begin{center}
		\includegraphics[width=0.485\textwidth]{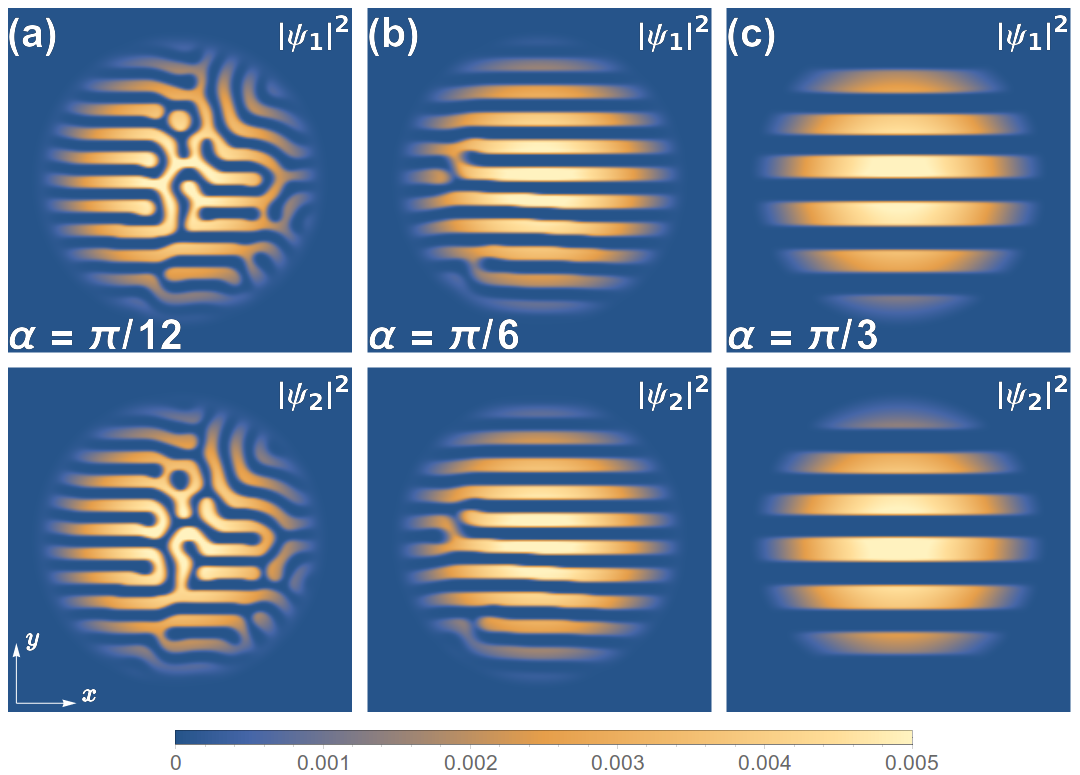}
	\end{center}\vspace{-0.5cm}
	\caption{\label{fig:anglen1} (Color online) Column density profiles $ |\psi_{1}|^{2} $ and $ |\psi_{2}|^{2} $ for stationary states with different tilted angles $ \alpha $. The other parameters are the same as in \reffig{fig:number}, except that $ N_{2} = 2 \times 10^6 $. The field of view is $ 55.66 \times 55.66 $ $ \mu{\rm m} $. The unit for the density plot is $ N_{1} a_{\perp}^{-2} $.}
\end{figure}

\begin{figure}[t]
	\begin{center}
		\includegraphics[width=0.485\textwidth]{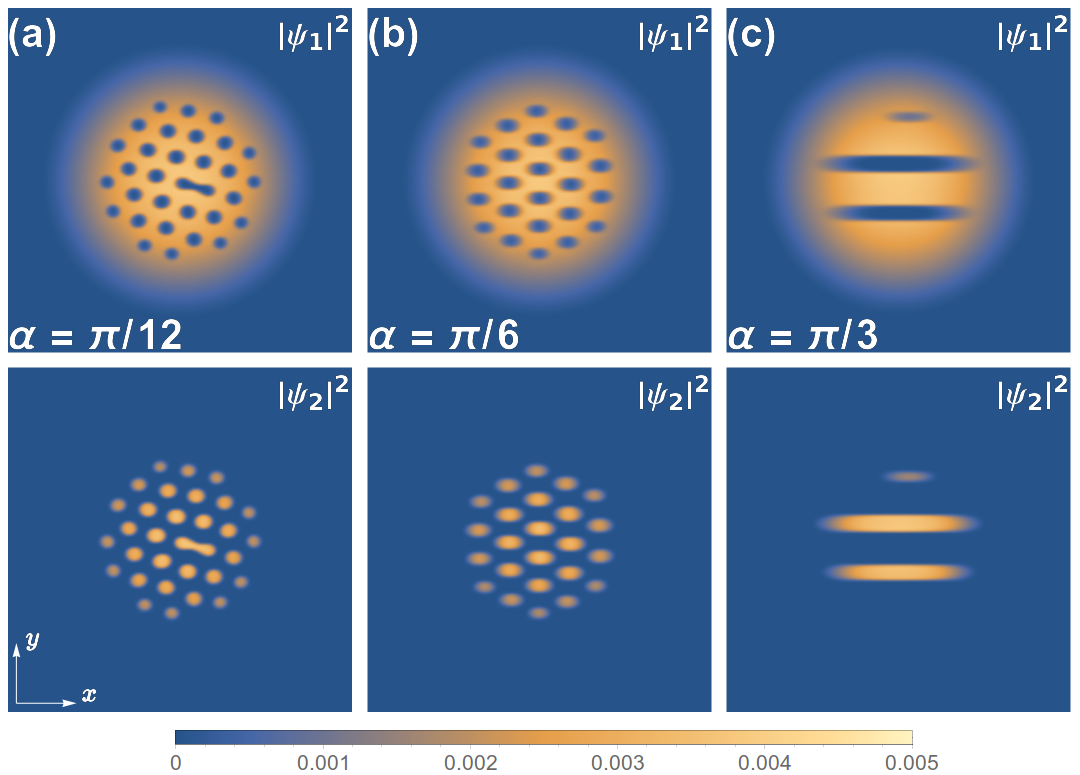}
	\end{center}\vspace{-0.5cm}
	\caption{\label{fig:anglen6} (Color online) Column density profiles $ |\psi_{1}|^{2} $ and $ |\psi_{2}|^{2} $ for stationary states with different tilted angles $ \alpha $. The other parameters are the same as in \reffig{fig:number}, except that $ N_{1} / N_{2} = 6 $. The field of view is $ 55.66 \times 55.66 $ $ \mu{\rm m} $. The unit for the density plot is $ N_{1} a_{\perp}^{-2} $.}
\end{figure}

The effect of a tilted polarization on the 2D dipolar Bose gas has recently been studied by mean-field theory \cite{tiltmean} and dynamic many-body theory \cite{tiltpimc}. We now investigate the effect of tilted polarization on the interface of a three-dimensional two-component dipolar BEC system. The parameters are taken to be $ a_{11} = a_{22} = a_{12} = 100 a_{B} $, $ \left( \omega_{\perp}, \omega_{z} \right) = 2 \pi \times \left( 100, 800 \right) $ Hz, and $ N_{1} = N_{2} = 2 \times 10^6 $. The angle $ \alpha $ of tilted polarization with respect to the $ z $ direction is increased from $ \pi / 12 $ to $ \pi / 3 $.

We plot the column density profiles in \reffig{fig:anglen1}. The results with $ \alpha = \pi / 12 $, $ \pi / 6 $, and $ \pi / 3 $ are shown. As the tilted angle $ \alpha $ increases, the filaments of the labyrinthine structure in \reffig{fig:number}(a) are gradually straightened, and the labyrinthine pattern is reorganized into a stripe phase, as shown in \reffig{fig:anglen1}(c), which is similar to the quantum phase transition in the 2D case in Ref. \cite{tiltpimc}. It is shown that the tilted polarization overcomes the effect of labyrinthine instability, and develops an ordered stripe pattern replacing the disordered labyrinthine pattern. 

We also modify the number ratio of atoms $ N_{1} / N_{2} = 6 $ in the two components. The column density profiles are plotted in \reffig{fig:anglen6} with $ \alpha = \pi / 12 $, $ \pi / 6 $ and $ \pi / 3 $. The droplet pattern induced by the large number ratio is also reshaped into a stripe pattern as the tilted polarization angle increases. It is shown that a crystallized droplet pattern occurs when $ \alpha = \pi / 6 $ and $ N_{1} / N_{2} = 6 $. Comparing \reffig{fig:anglen6}(c) with \reffig{fig:anglen1}(c), we find that the number of stripes is decreased as the number of atoms in component 2 is reduced. This is because there are insufficient atoms in component 2 to form many stripes.

\section{Conclusions}\label{sec:conclusion}

In conclusion, we have investigated the pattern formation of fingering instabilities and domain dynamics in an oppositely polarized two-component BEC with strong DDIs in three dimensions. We have demonstrated that dynamical fingering instabilities occur at the interface between two components of dipolar BECs, which breaks the rotational symmetry of the interface. Frog-shaped and mushroom-shaped patterns have been shown with different strengths of the DDIs. A Bogoliubov analysis gives a qualitative understanding of the interfacial instabilities of the two dipolar BECs, and a dispersion relation similar to that in classical fluids is obtained. Spontaneous density modulation and dipolar domain growth in the dynamics have also been demonstrated, in which we have analyzed the characteristic sizes of the dipolar domains corresponding to different patterns at the initial and later times in the evolution. We have also investigated the parameter dependence of the ground states, and found that the droplet patterns are formed due to the population imbalance in the two components.  Labyrinthine patterns grow as the trap ratio increases, and a striped phase appears as the angle of tilted polarization increases.

The findings in this and other recent works have revealed that including anisotropy and long-range dipole-dipole interactions results in rich physics of multicomponent BECs, with similarities to conventional viscous classical fluids. These findings shed light on the nature of dipolar BECs, which should be within reach of current experiments. For example, the system studied in this paper can be implemented by using the $ ^{7}S_{3} $ $ m_{j} = -3 $ and $ +3 $ states of $ ^{52} $Cr \cite{crbec2005,crbec2006}, the $ s $-wave scattering lengths can be modulated by Feshbach resonance \cite{lahaye2007}, and the strength of the DDI can be tuned by fast rotation of the magnetic field \cite{gamma}. In this paper, we ignore the dipolar relaxation and spinor dynamics. To observe the phenomena presented in this paper, the application of different dipolar species \cite{twospecies} may provide a more achievable approach. The experimental resolution of \textit{in situ} imaging is typically $ \sim \mu{\rm m} $, so the patterns of the BECs shown in the present paper can be measured. Recently, it was numerically shown that superfluid flow over a rough surface of a wire exhibits the properties of a boundary layer similar to those in classical fluids \cite{stagg2017}, which also establishes further connections between superfluids and classical fluids.  Such investigations are likely to be of relevance when examining superfluid flow in practical applications such as atomtronics and quantum metrology \cite{seaman2007,byrnes2015,moxley2016,lamico2017}.

\begin{acknowledgments}
K.-T.X. would like to thank E. Braaten for his kind support and comments. H.S. acknowledges support by JSPS KAKENHI Grants No. JP17K05595, No. JP16K05505, No. JP17K05596, and No. JP25103007. T.B. acknowledges support by the Shanghai Research Challenge Fund; New York University Global Seed Grants for Collaborative Research; National Natural Science Foundation of China (61571301); the Thousand Talents Program for Distinguished Young Scholars (D1210036A); the NSFC Research Fund for International Young Scientists (11650110425); NYU-ECNU Institute of Physics at NYU Shanghai; the Science and Technology Commission of Shanghai Municipality (17ZR1443600); and the China Science and Technology Exchange Center (NGA-16-001). 
\end{acknowledgments}



\end{document}